\documentclass[english]{IEEEtran}
\usepackage[T1]{fontenc}
\usepackage[latin9]{inputenc}
\usepackage{array}
\usepackage{multirow}
\usepackage{amsmath}
\usepackage{amsthm}
\usepackage{amssymb}
\usepackage{stackrel}
\usepackage{graphicx}
\usepackage{esint}
\PassOptionsToPackage{normalem}{ulem}
\usepackage{ulem}

\makeatletter

\providecommand{\tabularnewline}{\\}

\theoremstyle{plain}
\newtheorem{lem}{\protect\lemmaname}
\newenvironment{lyxlist}[1]
	{\begin{list}{}
		{\settowidth{\labelwidth}{#1}
		 \setlength{\leftmargin}{\labelwidth}
		 \addtolength{\leftmargin}{\labelsep}
		 }}
	{\end{list}}
\theoremstyle{definition}
\newtheorem{assumption}{Assumption}
\theoremstyle{remark}
\newtheorem{rem}{\protect\remarkname}
\theoremstyle{plain}
\newtheorem{thm}{\protect\theoremname}

\usepackage{cite}
\usepackage{multirow}
\usepackage{tikz}
\usetikzlibrary{decorations.pathmorphing, arrows.meta}

\makeatother

\usepackage{babel}
\providecommand{\lemmaname}{Lemma}
\providecommand{\remarkname}{Remark}
\providecommand{\theoremname}{Theorem}

\begin{document}
\title{LyLA-Therm: Lyapunov-based Langevin Adaptive Thermodynamic Neural
Network Controller}
\author{Saiedeh Akbari, Omkar Sudhir Patil, and Warren E. Dixon$^{*}$\thanks{$^{*}$Saiedeh Akbari, Omkar Sudhir Patil, and Warren E. Dixon are
with the department of Mechanical Engineering and Aerospace Engineering,
University of Florida, Gainesville, FL, 32611-6250 USA. Email: \{akbaris,
patilomkarsudhir, wdixon\}@ufl.edu. }\thanks{This research is supported in part by AFRL project FA8651-24-1-0018
and AFOSR grant FA9550-19-1-0169. Any opinions, findings, and conclusions
or recommendations expressed in this material are those of the author(s)
and do not necessarily reflect the views of the sponsoring agencies.}}

\maketitle
\global\long\def\SS{\mathbb{S}}%
\global\long\def\RR{\mathbb{R}}%
\global\long\def\EE{\mathbb{E}}%
\global\long\def\nz{\left\Vert z\right\Vert }%
\global\long\def\ne{\left\Vert e\right\Vert }%
\global\long\def\nt{\left\Vert \widetilde{\theta}\right\Vert }%
\global\long\def\nzz{\left\Vert z\right\Vert ^{2}}%
\global\long\def\nee{\left\Vert e\right\Vert ^{2}}%
\global\long\def\ntt{\left\Vert \widetilde{\theta}\right\Vert ^{2}}%
\global\long\def\tq{\triangleq}%
\global\long\def\Linf{\mathcal{L}_{\infty}}%
\global\long\def\yy{\mathcal{Y}}%
\global\long\def\uu{\mathcal{U}}%
\global\long\def\grad{\mathcal{\nabla_{\widetilde{\theta}}}}%
\global\long\def\tvec{\mathcal{\text{vec}}}%
\global\long\def\sgn{\mathcal{\text{sgn}}}%
\global\long\def\nsig{\left\Vert \Sigma\Sigma^{\top}\right\Vert _{\infty}}%
\global\long\def\pthat{\frac{\partial}{\partial\hat{\theta}}}%
\global\long\def\tr{\text{tr}}%
\global\long\def\proj{\text{proj}}%

\begin{abstract}
Thermodynamic principles can be employed to design parameter update
laws that address challenges such as the exploration vs. exploitation
dilemma. In this paper, inspired by the Langevin equation, an update
law is developed for a Lyapunov-based DNN control method, taking the
form of a stochastic differential equation. The drift term is designed
to minimize the system's generalized internal energy, while the diffusion
term is governed by a user-selected generalized temperature law, allowing
for more controlled fluctuations. The minimization of generalized
internal energy in this design fulfills the exploitation objective,
while the temperature-based stochastic noise ensures sufficient exploration.
Using a Lyapunov-based stability analysis, the proposed Lyapunov-based
Langevin Adaptive Thermodynamic (LyLA-Therm) neural network controller
achieves probabilistic convergence of the tracking and parameter estimation
errors to an ultimate bound. Simulation results demonstrate the effectiveness
of the proposed approach, with the LyLA-Therm architecture achieving
up to 20.66\% improvement in tracking errors, up to 20.89\% improvement
in function approximation errors, and up to 11.31\% improvement in
off-trajectory function approximation errors compared to the baseline
deterministic approach.
\end{abstract}


\section{Introduction\label{sec:Introduction}}

The process of parameter adaptation can be mathematically modeled
as the evolution of a state in a potentially high-dimensional space,
driven by gradients (forces) and subject to perturbations (stochastic
noise). This structure parallels the dynamics of physical systems
evolving on an energy landscape, where drift terms correspond to forces
derived from potential energy, and stochastic fluctuations resemble
thermal noise. Thermodynamics provides a well-established theoretical
framework for understanding how systems move toward equilibrium while
constantly experiencing fluctuations. Thermodynamic concepts such
as internal energy, entropy, and temperature laws offer mathematical
tools for governing the evolution of many systems.

Drawing inspiration from thermodynamics, innovative architectures
have been developed for control and learning. The works in \cite{Haddad.Chellaboina.ea2009}
and \cite{Haddad2019} examine the formulation of equilibrium and
non-equilibrium thermodynamics from a dynamical systems perspective.
Results such as \cite{Hui.Haddad2008,Hui.Haddad.ea2008a,Berg.Maithripala.ea2013,Haddad2021,Haddad.Chellaboina.ea2006,Haddad.Chellaboina.ea2007,Haddad.Hui.ea2007,Haddad.Lee2021,Haddad.Chahine2020}
use dynamical thermodynamics for addressing a variety of problems
such as distributed consensus and hybrid stabilization. Results such
as \cite{Sekimoto1997,Sekimoto1998,Sekimoto2010,Seifert2008,Seifert2012}
have developed theoretical frameworks for stochastic thermodynamics
by merging stochastic theory with non-equilibrium dynamics. Within
this context, thermodynamic state variables evolve according to an
over-damped Langevin equation, where fluctuation and dissipation forces
follow the Einstein relation (i.e., fluctuation-dissipation theorem
\cite{Einstein1905}), illustrating that diffusion results from both
thermal fluctuations and frictional dissipation \cite{Sekimoto1998,Sekimoto2010}.
Specifically, the over-damped condition means that the particle's
inertial effects are insignificant, so its motion is governed by an
instantaneous response to the random thermal forces (i.e., fluctuations)
and the viscous drag exerted by the medium (i.e., frictional dissipation).
The Einstein relation is fundamental here, as it establishes a direct
proportionality between the magnitude of these random thermal fluctuations,
the coefficient of frictional dissipation, and the system's absolute
temperature. Therefore, diffusion emerges directly from this interplay
such that particles are ceaselessly agitated by thermal energy, while
their resulting displacement is continuously shaped by dissipative
resistance of their environment. Consequently, Langevin equations
describe the balance between particle damping in the system and the
energy supplied to particles through thermal fluctuations. Despite
these theoretical developments of stochastic thermodynamics, their
application in control (especially adaptive control) is limited.

Regardless of whether a system is deterministic or stochastic, performing
system identification and control using Lyapunov-based deep neural
networks is challenged by conflicting goals of exploration and exploitation.
Exploitation requires the system to follow the reference trajectory
and achieve the control objectives. In adaptive control, the question
of exploration corresponds to whether the system takes a wide variety
of trajectories to gather sufficiently rich information and enhance
learning. Achieving sufficient exploration is challenged by dependence
of the deterministic states and the estimate of the parameters. Persistence
of excitation, finite excitation, and interval excitation are some
adaptive control methods to ensure sufficient information is gathered
for convergence of parameters \cite{Narendra.Annaswamy1987,Parikh.Kamalapurkar.ea2019,Pan.Yu2016,Pan.Zhang.ea2016,Cho.Shin.ea2017,Pan.Yu2018}.
Although it is a common practice in adaptive control to include a
dither noise signal for encouraging exploration, the design of such
noise is ad hoc and not constructive. In deterministic and stochastic
off-policy reinforcement learning algorithms, an uncorrelated Gaussian
noise is typically added to the action selection process to pursue
exploration \cite{Lillicrap.Hunt.ea2015,Fujimoto.Hoof.ea2018,Haarnoja.Zhou.ea2018,Abdolmaleki.Springenberg.ea2018}.
However, added noise is often designed heuristically, lacks a meaningful
and physical role, and may result in uncontrolled fluctuations that
destabilize the system. In this paper, we draw inspiration from the
Langevin equation to design a parameter update law that introduces
stochasticity into the update law as a means to encourage exploration.
A contribution of the work is that the stochasticity is constructively
designed based on a Lyapunov-based stability analysis.

Unlike existing approaches that rely solely on deterministic gradient
flows or introduce noise heuristically, this work proposes an adaptive
controller design where parameter evolution follows thermodynamic
laws. The update law is formulated as a Langevin-type stochastic differential
equation, with the drift term derived from energy minimization to
ensure exploitation and the diffusion term governed by a thermodynamic
temperature law to regulate stochastic fluctuations. This structured
approach provides a principled balance between exploration and exploitation.
As the system explores the parameter space, the temperature remains
high, promoting sufficient exploration. As tracking improves and errors
decrease, the temperature naturally decays, reducing fluctuations
and guiding the system toward stability. This thermodynamic inspired
adaptation allows for a smooth transition from exploration to exploitation
without introducing ad hoc noise.

Deep neural networks (DNNs) exploit the nested nonlinearities from
function composition to achieve enhanced function approximation capabilities
compared to shallow networks, as demonstrated empirically \cite{Rolnick.Tegmark2018,Liang.Srikant2016,Lamb.Bell.ea2023}.
Most established DNN approaches use offline learning with sampled
input-output datasets (e.g., \cite[Sec. 6.6]{Brunton.Kutz2019} and
\cite{Noda.Arie.ea2014,Sarikaya.Corso.ea2017,Nguyen.Cheah2022}).
These DNNs function as feedforward components with fixed weights that
cannot be adapted in real-time through Lyapunov stability-based updates,
resulting in open-loop configurations lacking formal stability assurances.
Recent advances address this limitation by developing Lyapunov-based
DNNs (Lb-DNN) that employ Lyapunov-driven methodologies for real-time
weight adaptation \cite{Patil.Le.ea2022,Patil.Le.ea.2022,Griffis.Patil.ea.2023,Akbari.Griffis.ea2023,Akbari.Nino.ea2024}.
Specifically, Lb-DNN weights are modified through analytical update
rules derived from Lyapunov stability theory, enabling online adaptation
without requiring offline pre-training.

In this paper, a novel Langevin type Lyapunov-based deep neural network
update law is introduced (i.e., Lyapunov-based Langevin Adaptive Thermodynamic
(LyLA-Therm) neural network controller). The key contributions of
this work are:
\begin{enumerate}
\item Thermodynamic-inspired update law: the update law is formulated as
a stochastic differential equation, where the drift term minimizes
generalized internal energy to drive exploitation, and the diffusion
term, controlled by a user-defined generalized temperature law, introduces
stochasticity for exploration. The generalized temperature law adapts
based on tracking errors and parameter estimates, ensuring a natural
decay of exploration as learning progresses.
\item Controller design: an adaptive controller is designed to assist the
exploitation objective while accounting for the cost of exploration
by compensating for part of the stochasticity in the update law.
\item Stability analysis: a constructive Lyapunov-based stability analysis
is conducted to guarantee the probabilistic uniform ultimate boundedness
of tracking and parameter estimation errors, while accounting for
the flexibility of user-selected generalized temperature laws.
\item The proposed LyLA-Therm method is validated through comprehensive
simulations on a five-dimensional nonlinear dynamical system using
three distinct generalized temperature law formulations. Results demonstrate
19.97\%--20.66\% improvement in tracking errors, 20.44\%--20.89\%
improvement in function approximation errors, and 6.64\%--11.31\%
improvement in off-trajectory function approximation errors compared
to baseline deterministic approaches, validating the effectiveness
of different temperature law strategies while maintaining probabilistic
stability guarantees.
\end{enumerate}

\section{Preliminaries}

Let $\mathbf{0}_{m\times n}$ and $I_{m\times n}$ denote the $m\times n$
dimensional zero and identity matrices, respectively. The right pseudo
inverse of a full-row-rank matrix $A$ is defined as $A^{+}\left(\cdot\right)\tq A^{\top}\left(\cdot\right)\left(A\left(\cdot\right)A^{\top}\left(\cdot\right)\right)^{-1}$.
The expected value is given by ${\rm E}\left[X\right]=\intop_{-\infty}^{\infty}x\cdot\mathbf{F}\left(x\right){\rm d}x$,
where $\mathbf{F}\left(x\right)$ is the probability density function
of the continuous random variable $X$. For a matrix $A\tq\left[a_{i,j}\right]\in\RR^{n\times m}$,
where $a_{i,j}$ is the element on the $i^{\text{th}}$ row of the
$j^{\text{th}}$ column of the matrix, the vectorization operator
is defined as $\tvec\left(A\right)=\left[a_{1,1},\ldots,a_{n,1},\ldots,a_{1,m},\ldots,a_{n,m}\right]^{\top}\in\RR^{nm}$.
For a square matrix $A\in\RR^{n\times n}$, the trace operator is
defined as $\text{tr}\left(A\right)=\stackrel[i=1]{n}{\sum}a_{i,i}$,
where $a_{i,i}$ represents the element on the $i^{\text{th}}$ row
of the $i^{\text{th}}$ column. From \cite[Chapter 1, Eq. 25]{Magnus.Neudecker2019},
\begin{equation}
\text{tr}\left(AB\right)=\text{tr}\left(BA\right).\label{eq: order of multiplication property}
\end{equation}
 The right-to-left matrix product operator is represented by $\stackrel{\curvearrowleft}{\prod}$,
i.e., $\stackrel{\curvearrowleft}{\stackrel[p=1]{m}{\prod}}A_{p}=A_{m}\ldots A_{2}A_{1}$
and $\stackrel{\curvearrowleft}{\stackrel[p=a]{m}{\prod}}A_{p}=I$
if $a>m$. The Kronecker product is denoted by $\otimes$. From \cite{Patil.Le.ea.2022}
and given any $A\in\mathbb{R}^{p\times a}$, $B\in\mathbb{R}^{a\times r}$,
and $C\in\mathbb{R}^{r\times s}$, differentiating $\mathrm{vec}\left(ABC\right)$
on both sides with respect to $\mathrm{vec}\left(B\right)$ yields
\begin{eqnarray}
\frac{\partial}{\partial\mathrm{vec}\left(B\right)}\mathrm{vec}\left(ABC\right) & = & C^{\top}\otimes A.\label{eq:vec_diff_prop}
\end{eqnarray}
The projection operator is given by \cite[Eq. E.4]{Krstic.Kanellakopoulos.ea1995}
{\footnotesize{}
\begin{equation}
\proj\left(\mathfrak{m}\right)=\begin{cases}
\mathfrak{m}, & \hat{\theta}\in\overset{\circ}{\Pi}\text{ or }\nabla\mathcal{P}^{\top}\mathfrak{m}\leq0,\\
\left(1-\mathfrak{h}\left(\hat{\theta}\right)\frac{\nabla\mathcal{P}\nabla\mathcal{P}^{\top}}{\left\Vert \nabla\mathcal{P}\right\Vert ^{2}}\right)\mathfrak{m}, & \hat{\theta}\in\Pi_{\varepsilon}\backslash\overset{\circ}{\Pi}\text{ and }\nabla\mathcal{P}^{\top}\mathfrak{m}>0,
\end{cases}\label{eq: projection definition}
\end{equation}
}where $\mathcal{P}:\RR^{p}\to\RR$ denotes a smooth convex function,
$\overset{\circ}{\Pi}$ denotes the interior of $\Pi=\left\{ \hat{\theta}\in\RR^{p}:\mathcal{P}\left(\hat{\theta}\right)\leq0\right\} $,
the convex set $\Pi_{\epsilon}=\left\{ \hat{\theta}\in\RR^{p}:\mathcal{P}\left(\hat{\theta}\right)\leq\epsilon\right\} $
is a union of the set $\Pi$ and an $\mathcal{O}\left(\epsilon\right)$-boundary
layer around it, and $\mathfrak{h}\left(\hat{\theta}\right)=\min\left(1,\frac{1}{\epsilon}\mathcal{P}\left(\hat{\theta}\right)\right)$.
The $a$-norm is denoted by $\left\Vert \cdot\right\Vert _{a}$, where
the subscript is suppressed when $p=2$. The Frobenius norm is denoted
by $\left\Vert \cdot\right\Vert _{F}\triangleq\left\Vert \mathrm{vec}(\cdot)\right\Vert $.
The space of $k$-times differentiable functions is denoted by $\mathcal{C}^{k}$,
and a $\mathcal{C}^{\infty}$-smooth function is an infinitely differentiable
function. In the filtered probability space of $\left(\mho,\,\mathbb{F},\,\mathbb{F}_{t},{\rm P}\right)$,
$\mho$ represents the event space, $\mathbb{F}$ denotes a $\sigma$-algebra
of the subsets of $\mho$, $\mathbb{F}_{t}$ is a complete filtration
given by the family of $\sigma$-algebras up to time $t$, i.e., $\mathbb{F}_{S}:\mathbb{F}_{S}\subseteq\mathbb{F}_{t}\,\forall t\in\left[0,t\right]$,
and ${\rm P}$ is a probability measure, where the filtration is complete
in the sense it includes all events with probability measure zero
(see \cite{Lanchares.Haddad2023}). Consider a probability space of
$\left(\mho,\,\mathbb{F},\,{\rm P}\right)$. Then, for any events
$A,\,B\in\mathbb{F}$ such that $A\subseteq B$, the monotonicity
property states that \cite[eq. 2.5]{Billingsley2017} 
\begin{equation}
{\rm P}\left(A\right)\leq{\rm P}\left(B\right).\label{eq:monotonicity}
\end{equation}

Consider a stochastic differential equation (SDE) as ${\rm d}x={\bf f}\left(x\right){\rm d}t+{\bf g}\left(x,t\right){\rm d}\omega$.
Then, for some function $V\in\mathcal{C}^{2}$ associated with this
SDE, let the infinitesimal generator $\mathcal{L}$ of the function
$V\left(x\right)$ be defined as \cite[eq. 4.12]{Kushner1967}
\begin{gather}
\mathcal{L}V\tq\frac{\partial V}{\partial x}{\bf f}\left(x\right)+\frac{1}{2}\text{tr}\left({\bf g}\left(x,t\right)^{\top}\frac{\partial^{2}V}{\partial x^{2}}{\bf g}\left(x,t\right)\right).\label{eq: LV_L-1}
\end{gather}

\begin{lem}
\label{thm:probability}\textup{ (\cite[Lemma 1]{Akbari.Nino.ea2024})
For the Ito process ${\tt z}\in\RR^{n}$ and function ${\tt V}$,
assume }
\end{lem}
\begin{lyxlist}{00.00.0000}
\item [{(A1)}] \label{(A1)--is}${\tt V}$ is non-negative, ${\tt V}\left(0\right)=0$,
and ${\tt V}\in\mathcal{C}^{2}$ over the open and connected set $Q_{m}\tq\left\{ {\tt z}:{\tt V}\left({\tt z}\right)<m\right\} $,
where $m\in\RR_{>0}$ is a bounding constant,
\item [{(A2)}] \label{(A2)--is}${\tt z}\left(t\right)$ is a continuous
strong Markov process defined until at least some $\tau^{\prime}>\tau_{m}=\inf\left\{ t:{\tt z}\left(t\right)\notin Q_{m}\right\} $
with probability one.\footnote{This assumption guarantees the existence of the process up to $\tau^{\prime}$
with probability one.}
\end{lyxlist}
If $\mathcal{L}{\tt V}\left({\tt z}\right)\leq-{\tt k}_{1}{\tt V}\left({\tt z}\right)+{\tt k}_{2}$
in $Q_{m}$ for ${\tt k}_{1},{\tt k}_{2}>0$, then for $\ell\leq m$,
${\tt z}\left(t\right)$ is uniformly ultimately bounded in probability
(UUB-p) with the probability
\begin{align*}
{\rm P}\left(\underset{t\leq s<\infty}{\sup}{\tt V}\left({\tt z}\left(s\right)\right)\geq\ell\right) & \leq\frac{1}{m}{\tt V}\left({\tt z}\left(0\right)\right)\\
 & \hspace{1em}+\frac{1}{\ell}{\tt V}\left({\tt z}\left(0\right)\right)\exp\left(-{\tt k}_{1}t\right)+\frac{{\tt k}_{2}}{{\tt k}_{1}\ell}.
\end{align*}

\subsection{Deep Neural Network Model}

Let $\kappa\in\mathbb{R}^{L_{0}}$ denote the DNN input, and $\theta\in\mathbb{R}^{p}$
denote the vector of DNN parameters (i.e., weights and bias terms).
A fully-connected feedforward DNN $\Phi(\kappa,\theta)$ with $k\in\mathbb{Z}_{>0}$
hidden layers and output size $L_{k+1}\in\mathbb{Z}_{>0}$ is defined
using a recursive relation $\varphi_{j}\in\mathbb{R}^{L_{j+1}}$ modeled
as \cite{Patil.Le.ea2022}
\begin{eqnarray}
\varphi_{j} & \triangleq & \begin{cases}
V_{j+1}^{\top}\kappa_{a}, & j=0,\\
V_{j+1}^{\top}\phi_{j}\left(\varphi_{j-1}\right) & j\in\left\{ 1,\ldots,k\right\} ,
\end{cases}\label{eq:DNN}
\end{eqnarray}
where $\Phi(\kappa,\theta)=\varphi_{k}$ , $\kappa_{a}\tq\left[\kappa^{\top},1\right]^{\top}$
denotes the augmented input that accounts or the bias terms, $L_{j}\in\mathbb{Z}_{>0}$
denotes the number of neurons in the $j^{\textrm{th}}$ layer with
$L_{j}^{a}\tq L_{j}+1$, and $V_{j+1}\in\mathbb{R}^{L_{j}^{a}\times L_{j+1}}$
denotes the matrix of weights and biases, for all $j\in\left\{ 0,\ldots,k\right\} $.

The vector of activation functions is denoted by $\phi_{j}:\mathbb{R}^{L_{j}}\to\mathbb{R}^{L_{j}^{a}}$
for all $j\in\left\{ 1,\ldots,k\right\} $. The vector of activation
functions can be composed of various activation functions, and hence,
may be represented as $\phi_{j}=\left[\varsigma_{1},\ldots,\varsigma_{L_{j}},1\right]^{\top}$
for all $j\in\left\{ 1,\ldots,k\right\} $, where $\varsigma_{j}:\mathbb{R}\to\mathbb{R}$
for all $j\in\left\{ 1,\ldots,L_{j}\right\} $ denotes a piece-wise
continuously differentiable activation function, where 1 denotes the
augmented hidden layer that accounts for the bias terms. For the DNN
architecture in (\ref{eq:DNN}), the vector of DNN weights is $\theta\triangleq\left[\mathrm{vec}\left(V_{1}\right){}^{\top},\ldots,\mathrm{vec}\left(V_{k}\right){}^{\top}\right]^{\top}$
with size $p=\sum_{j=0}^{k}L_{j}^{a}L_{j+1}$.

Consider $y_{j}\in\mathbb{R}^{L_{j}}$ where $y_{j}=\left[y_{1},\ldots,y_{L_{j}}\right]$
with $y_{i}\in\mathbb{R}$ for all $i\in\left\{ 1,\ldots,L_{j}\right\} $.
The Jacobian $\frac{\partial\phi_{j}}{\partial y_{j}}:\mathbb{R}^{L_{j}}\to\mathbb{R}^{L_{j}^{a}\times L_{j}}$
of the activation function vector at the $j^{\mathrm{th}}$ layer
is given by $\left[\varsigma_{1}^{\prime}(y_{1})\eta_{1},\ldots,\varsigma_{L_{j}}^{\prime}(y_{L_{j}})\eta_{L_{j}},\mathbf{0}_{L_{j}}\right]^{\top}\in\mathbb{R}^{L_{j}^{a}\times L_{j}}$,
where $\varsigma_{j}^{\prime}$ denotes the derivative of $\varsigma_{j}$
with respect to its argument for $j\in\left\{ 1,\ldots,L_{j}\right\} $,
$\eta_{i}$ is the $i^{\text{th}}$ standard basis vector in $\mathbb{R}^{L_{j}}$,
and $\mathbf{0}_{L_{j}}$ is the zero vector in $\mathbb{R}_{L_{j}}$.

Let the gradient of the DNN with respect to the weights be denoted
by $\Phi^{\prime}(\kappa,\theta)\triangleq\frac{\partial}{\partial\theta}\Phi(\kappa,\theta)$,
which can be represented as $\Phi^{\prime}(\kappa,\theta)=\left[\frac{\partial}{\partial{\rm vec}(V_{1})}\Phi(\kappa,\theta),\ldots,\frac{\partial}{\partial{\rm vec}(V_{k+1})}\Phi(\kappa,\theta)\right]\in\mathbb{R}^{L_{k+1}\times p}$,
where $\frac{\partial}{\partial{\rm vec}\left(V_{j}\right)}\Phi\left(\kappa,\theta\right)\in\mathbb{R}^{L_{k+1}\times L_{j-1}^{a}L_{j}}$
for all $j\in\left\{ 1,\ldots,k+1\right\} $. Then, using (\ref{eq:DNN})
and the property of the vectorization operator in (\ref{eq:vec_diff_prop})
yields
\begin{align}
\Phi^{\prime}(\kappa,\theta) & =\left(\stackrel{\curvearrowleft}{\stackrel[\ell=j+1]{k}{\prod}}V_{\ell+1}^{\top}\frac{\partial\phi_{\ell}}{\partial\varphi_{\ell-1}}\right)\left(I_{L_{j+1}}\otimes\varrho_{j}\right),\label{eq:DNN Gradient}
\end{align}
for $j\in\left\{ 0,\ldots,k\right\} $, where $\varrho_{j}=\kappa_{a}^{\top}$
if $j=0$ and $\varrho_{j}=\phi_{j}^{\top}\left(\varphi_{j-1}\right)$
if $j\in\left\{ 1,\ldots,k\right\} $.
\begin{assumption}
\label{assm:activation bounds} For each $j\in\left\{ 0,\ldots,k\right\} $,
the activation function $\phi_{j}$, its Jacobian $\phi_{j}^{\prime}$,
and Hessian $\phi_{j}^{\prime\prime}\left(y\right)\triangleq\frac{\partial^{2}}{\partial y^{2}}\phi_{j}\left(y\right)$
are bounded as
\begin{eqnarray}
\left\Vert \phi_{j}\left(y\right)\right\Vert  & \leq & \mathfrak{a}_{1}\left\Vert y\right\Vert +\mathfrak{a}_{0},\nonumber \\
\left\Vert \phi_{j}^{\prime}\left(y\right)\right\Vert  & \leq & \mathfrak{b}_{0},\nonumber \\
\left\Vert \phi_{j}^{\prime\prime}\left(y\right)\right\Vert  & \leq & \mathfrak{c}_{0},\label{eq:Activation Bounds}
\end{eqnarray}
where $\mathfrak{a}_{0},\mathfrak{a}_{1},\mathfrak{b}_{0},\mathfrak{c}_{0}\in\mathbb{R}_{\geq0}$
are known constants.
\end{assumption}
\begin{rem}
\label{rem:activation bounds} Most activation functions used in practice
satisfy Assumption \ref{assm:activation bounds}. Specifically, sigmoidal
activation functions (e.g., logistic function, hyperbolic tangent
etc.) have $\left\Vert \phi_{j}\left(y\right)\right\Vert $, $\left\Vert \phi_{j}^{\prime}\left(y\right)\right\Vert $,
and $\left\Vert \phi_{j}^{\prime\prime}\left(y\right)\right\Vert $
bounded uniformly by constants. Smooth approximations of rectified
linear unit (ReLUs) such as Swish grow linearly, and hence satisfy
the bound $\left\Vert \phi_{j}\left(y\right)\right\Vert \leq\mathfrak{a}_{1}\left\Vert y\right\Vert +\mathfrak{a}_{0}$
of Assumption \ref{assm:activation bounds}.
\end{rem}

\section{Dynamics and Control Objective}

Consider a control-affine nonlinear dynamical system as
\begin{equation}
\dot{x}=f\left(x\right)+g\left(x\right)u\left(t\right),\label{eq: dynamical system}
\end{equation}
where $t\in\RR_{\geq0}$ denotes time, $x\in\RR^{n}$ denotes the
measurable state variable, $f:\RR^{n}\to\RR^{n}$ denotes an unknown
continuous drift vector field, $g:\RR^{n}\to\RR^{n\times r}$ denotes
the known full row rank and bounded control effectiveness matrix,
and $u\in\RR^{r}$ denotes the control input. 

The control objective is to design an adaptive controller such that
the state $x$ converges (in expectation) to a $\mathcal{C}^{2}$-smooth
user-defined desired trajectory $x_{d}:\RR_{\geq0}\to\RR^{n}$. To
quantify the control objective, the tracking error $e\in\RR^{n}$
is defined as
\begin{equation}
e\tq x-x_{d}.\label{eq: tracking error}
\end{equation}

\begin{assumption}
\label{thm: boundedness of desired trajectory}There exists known
constants $\overline{x}_{d},\overline{\dot{x}}_{d}\in\RR_{>0}$ such
that $\left\Vert x_{d}\right\Vert \leq\overline{x}_{d}$ and $\left\Vert \dot{x}_{d}\right\Vert \leq\overline{\dot{x}}_{d}$.
\end{assumption}

\section{Control Design}

\subsection{Feedforward DNN Approximation\label{subsec:Thermodynamic-Neural-Network}}

Let $\Omega\subset\RR^{n}$ denote the compact set over which the
universal function approximation property holds. The DNN-based approximation
of $f\left(x\right)$ is given by $\hat{f}=\Phi\left(x,\hat{\theta}\right)$,
where $\Phi:\RR^{n}\times\RR^{p}\to\RR^{n}$ denote a general DNN
architecture, and $\hat{\theta}\in\RR^{p}$ denotes the DNN parameter
estimates that are subsequently designed. The unknown drift vector
in (\ref{eq: dynamical system}) can be modeled using a DNN as
\begin{equation}
f\left(x\right)=\Phi\left(x,\theta^{*}\right)+\varepsilon\left(x\right),\label{eq: Lb-ThNN model}
\end{equation}
where $\varepsilon:\RR^{n}\to\RR^{n}$ is an unknown function representing
the reconstruction error that is bounded by an unknown $\overline{\varepsilon}\in\RR_{>0}$
as
\begin{equation}
\underset{x\in\Omega}{\sup}\left\Vert \varepsilon\left(x\right)\right\Vert \leq\overline{\varepsilon}.\label{eq: varepsilon bound}
\end{equation}

The approximation objective is to determine optimal estimates of $\theta$
such that $x\mapsto\Phi\left(x,\hat{\theta}\right)$ approximates
$x\mapsto f\left(x\right)$ with minimal error for any $x\in\Omega$.
To quantify the approximation objective, let the loss function $\mathfrak{L}:\RR^{p}\to\RR_{\geq0}$
b defined as
\begin{equation}
\mathfrak{L}\left(\theta\right)\tq\int_{\Omega}\left(\left\Vert f\left(x\right)-\Phi\left(x,\theta\right)\right\Vert ^{2}+\mathfrak{s}\left\Vert \theta\right\Vert ^{2}\right){\rm d}\mathbf{m}\left(x\right),\label{eq: loss function}
\end{equation}
where $\mathbf{m}$ denotes the Lebesgue measure, $\mathfrak{s}\in\RR_{>0}$
denotes a regularizing constant, and the term $\mathfrak{s}\left\Vert \theta\right\Vert ^{2}$
represents $L_{2}$ regularization. Let $\mathbf{\Omega}\subset\RR^{p}$
denote a user-selected compact and convex parameter search space with
a smooth boundary, satisfying $\mathbf{0}_{p}\in\text{int}\left(\mathbf{\Omega}\right)$.
Additionally, define $\overline{\theta}\tq\underset{\theta\in\mathbf{\Omega}}{\max}\left\Vert \theta\right\Vert $
to be a bound on the user-selected search space. For instance, if
${\bf \Omega}$ is constructed as a ball of radius 5, then $\overline{\theta}=5$.
The objective is to identify the vector of ideal DNN parameters $\theta^{*}\in\mathbf{\Omega}$
defined as
\begin{equation}
\theta^{*}\tq\underset{\theta\in\mathbf{\Omega}}{\arg\,\min}\mathfrak{L}\left(\theta\right).\label{eq: ideal weights}
\end{equation}
For the solution of (\ref{eq: ideal weights}) to be unique, the loss
function in (\ref{eq: loss function}) is required to be strictly
convex over the set $\mathbf{\Omega}$. Therefore, since the regularizing
term $\mathfrak{s}\left\Vert \theta\right\Vert ^{2}$ in (\ref{eq: loss function})
is strictly convex, there exists $\mathfrak{s}$ which ensures $\mathfrak{L}$$\left(\theta\right)$
is convex for all $\theta\in\mathbf{\Omega}$. Selecting high values
of $\mathfrak{s}$ can obscure the contribution of the $\left\Vert f\left(x\right)-\Phi\left(x,\theta\right)\right\Vert ^{2}$
term in the loss function as well as causing underfitting \cite[Sec. 7.1.1]{Goodfellow.Bengio.ea2016}.
For less restrictive identifiability conditions under which strict
convexity is guaranteed, refer to \cite[Lem. 1]{Hart.Patil.ea2025}.

\subsection{Adaptive Thermodynamic Update Law}

To address the question of exploration versus exploitation in DNNs,
inspiration is drawn from Langevin dynamics to design a stochastic
differential adaptive update law as
\begin{equation}
{\rm d}\hat{\theta}\tq\proj\left(\varrho\left(x,\hat{\theta},t\right)\right){\rm d}t+\proj\left(\varsigma\left(x,\hat{\theta},t\right){\rm d}\omega\right),\label{eq: Thermodynamic update law}
\end{equation}
where $\varrho:\RR^{n}\times\RR^{p}\times\RR_{\geq0}\to\RR^{p}$ denotes
the drift process and $\varsigma:\RR^{n}\times\RR^{p}\times\RR_{\geq0}\to\RR^{p\times p}$
denotes the diffusion term, and $\omega\in\mathbb{R}^{p}$ denotes
Wiener process. Based on following Lemma, the projection operator
is used to ensure $\hat{\theta}$ is bounded as $\left\Vert \hat{\theta}\right\Vert \leq\overline{\theta}$. 
\begin{lem}
\label{lem: boundedness of theta hat}Let $\hat{\theta}\left(0\right)\in\Pi_{\epsilon}$.
The projection operators in (\ref{eq: Thermodynamic update law})
ensure that the solution $\hat{\theta}\left(t\right)$ remains in
$\Pi_{\epsilon}$.
\end{lem}
\begin{IEEEproof}
From \cite[Page 513]{Krstic.Kanellakopoulos.ea1995}, it is known
that for any input $\mathfrak{m}$ to the projection operator, $\nabla_{\hat{\theta}}\mathcal{P}^{\top}\text{proj}\left(\mathfrak{m}\right)\leq0$,
whenever $\hat{\theta}\in\partial\Pi_{\epsilon}$, meaning $\hat{\theta}$
is at the boundary. This property ensures that both drift increment
and the diffusion terms are either tangent to or directed inward relative
to $\Pi_{\epsilon}$. At any point $\hat{\theta}\in\partial\Pi_{\epsilon}$,
the drift increment $\proj\left(\varrho\left(x,\hat{\theta}\right)\right)$
satisfies $\nabla_{\hat{\theta}}\mathcal{P}^{\top}\text{proj}\left(\varrho\left(x,\hat{\theta}\right)\right)\leq0$.
Similarly, the diffusion term $\proj\left(\varsigma\left(x,\hat{\theta}\right){\rm d}\omega\right)$
also satisfies $\nabla_{\hat{\theta}}\mathcal{P}^{\top}\proj\left(\varsigma\left(x,\hat{\theta}\right){\rm d}\omega\right)\leq0$.
Therefore, at the boundary $\partial\Pi_{\epsilon}$, both drift and
diffusion terms are either tangent or directed inward. Since $\hat{\theta}\left(0\right)\in\Pi_{\epsilon}$,
it follows that $\hat{\theta}\left(t\right)\in\Pi_{\epsilon},$ for
all $t\in\RR_{\geq0}$. 
\end{IEEEproof}
A thermodynamic approach is used for designing the drift and diffusion
processes. In the thermodynamic framework, the exploitation objective
can be described in terms of minimizing the generalized internal energy
loss function $U:\RR^{n}\times\RR^{p}\times\RR_{\geq0}\to\RR$ defined
as \cite[Sec. 2.4.2]{Fradkov2007}
\begin{align}
U\left(x,\hat{\theta},t\right) & \tq\frac{1}{2}\frac{{\rm d}}{{\rm d}t}e^{\top}e+\frac{1}{2}\sigma\hat{\theta}^{\top}\hat{\theta}\nonumber \\
 & =e^{\top}\dot{e}+\frac{1}{2}\sigma\hat{\theta}^{\top}\hat{\theta},\label{eq: internal energy}
\end{align}
where $\sigma\in\RR_{>0}$ is a user-defined forgetting factor. By
the definition of generalized internal energy in (\ref{eq: internal energy}),
the drift process is defined as
\begin{equation}
\varrho\left(x,\hat{\theta},t\right)\tq-\frac{\partial}{\partial\hat{\theta}}U\left(x,\hat{\theta},t\right).\label{eq: drift process - partial form}
\end{equation}
Inspired from the Langevin equation, the diffusion process is defined
as
\begin{equation}
\varsigma\left(x,\hat{\theta},t\right)\tq\sqrt{k_{T}T\left(x,\hat{\theta},t\right)},\label{eq: general form of varsigma}
\end{equation}
where $k_{T}\in\RR_{>0}$ is a constant gain, and $T:\RR^{n}\times\RR^{p}\times\RR_{\geq0}\to\RR$
denotes the generalized temperature law defined as
\begin{equation}
T\left(x,\hat{\theta},t\right)\tq e^{\top}\mu\left(x,\hat{\theta},t\right),\label{eq: Temperature law}
\end{equation}
where $\mu:\RR^{n}\times\RR^{p}\times\RR_{\geq0}\to\RR^{n}$ is a
user-selected term. While different designs of $\mu$ may be possible,
based on the subsequent stability analysis, this result requires the
term $\mu$ to be designed such that the generalized temperature law
in (\ref{eq: Temperature law}) is positive and differentiable and
that 
\begin{equation}
\left\Vert \frac{\partial\mu}{\partial\hat{\theta}}\right\Vert \leq{\tt c}_{2}\left\Vert e\right\Vert +{\tt c}_{1}\left\Vert \hat{\theta}\right\Vert +{\tt c}_{0}.\label{eq: bound on the partial of mu}
\end{equation}

Consistent with the Langevin equation framework and the fluctuation-dissipation
theorem discussed in Section \ref{sec:Introduction}, the diffusion
term in (\ref{eq: general form of varsigma}) is designed to represent
the intensity of stochastic perturbations influencing the system.
This formulation, particularly due to its dependence on the square
root of generalized temperature term $T\left(x,\hat{\theta},t\right)$,
reflects the characteristic scaling of thermal noise magnitude observed
in physical systems. Thus, $T\left(x,\hat{\theta},t\right)$ effectively
parameterizes the magnitude of thermal fluctuations, and the overall
diffusion term $\varsigma\left(x,\hat{\theta},t\right)$ embodies
the crucial interplay between this random agitation and the system's
inherent frictional dissipation, as conceptualized by the fluctuation-dissipation
theorem.

Substituting (\ref{eq: Temperature law}) into (\ref{eq: general form of varsigma})
yields the diffusion law as
\begin{align}
\varsigma\left(x,\hat{\theta},t\right) & \tq\sqrt{k_{T}e^{\top}\mu\left(x,\hat{\theta},t\right)}.\label{eq: diffusion term}
\end{align}

The following assumptions are introduced to assist with the design
of the update law based on the stability analysis in Section \ref{sec:Stability-Analysis}.

\subsection{Controller and Closed-Loop Error System}

To compensate for the uncertainties that appear in the stability analysis,
the developed LyLA-Therm adaptive update law is incorporated into
the controller designed as
\begin{align}
u & \tq g^{+}\left(x\right)\bigg(\dot{x}_{d}-k_{e}e-\Phi\left(x,\hat{\theta}\right),\nonumber \\
 & \hspace{1em}-\frac{1}{2}\left(p+1\right)\gamma k_{T}\mu\bigg)\label{eq: controller}
\end{align}
where $k_{e},\gamma\in\RR_{>0}$ are user-selected constant gains,
and $p$ was previously introduced to denote the number of the LyLA-Therm
weights. Substituting (\ref{eq: dynamical system}) and (\ref{eq: controller})
into the time derivative of the tracking error in (\ref{eq: tracking error})
and canceling out the cross terms yields{\small{}
\begin{equation}
\dot{e}=f\left(x\right)-k_{e}e-\Phi\left(x,\hat{\theta}\right)-\frac{1}{2}\left(p+1\right)\gamma k_{T}\mu.\label{eq: CLES with f(x)}
\end{equation}
}Substituting (\ref{eq: Lb-ThNN model}) into (\ref{eq: CLES with f(x)})
yields 
\begin{align}
\dot{e} & =\Phi\left(x,\theta^{*}\right)-\Phi\left(x,\hat{\theta}\right)-k_{e}e\nonumber \\
 & \hspace{1em}-\frac{1}{2}\left(p+1\right)\gamma k_{T}\mu+\varepsilon\left(x\right).\label{eq: CLES with f(x) - 1}
\end{align}
Let the corresponding estimation error be defined as
\begin{equation}
\tilde{\theta}\tq\theta^{*}-\hat{\theta}.\label{eq: Estimation error}
\end{equation}
To facilitate the subsequent stability analysis, a first-order Taylor
approximation (cf., \cite[Eq. 22]{Patil.Le.ea.2022} and \cite{Lewis.Yesildirek.ea1996})
is used on $\Phi\left(x,\theta^{*}\right)-\Phi\left(x,\hat{\theta}\right)$
to yield
\begin{equation}
\Phi\left(x,\theta^{*}\right)-\Phi\left(x,\hat{\theta}\right)=\Phi^{\prime}\left(x,\hat{\theta}\right)\tilde{\theta}+\mathcal{R}\left(x,\tilde{\theta}\right),\label{eq: Taylor approximation}
\end{equation}
where $\mathcal{R}:\RR^{n}\times\RR^{p}\to\RR^{n}$ denotes the Lagrange
remainder term. The Lagrange remainder in (\ref{eq: Taylor approximation})
is unknown; however, following lemma provides a polynomial bound for
the term.
\begin{lem}
\label{lem: Lagrange remainder bound} There exists a polynomial function
$\rho_{0}:\RR_{\geq0}\to\RR_{\geq0}$ of the form $\rho_{0}\left(\left\Vert x\right\Vert \right)={\tt a}_{2}\left\Vert x\right\Vert ^{2}+{\tt a}_{1}\left\Vert x\right\Vert +{\tt a}_{0}$
with some known constants ${\tt a}_{0},{\tt a}_{1},{\tt a}_{2}\in\RR_{>0}$
such that the Lagrange remainder term can be bounded as $\left\Vert \mathcal{R}\left(x,\tilde{\theta}\right)\right\Vert \leq\rho_{0}\left(\left\Vert x\right\Vert \right)\left\Vert \tilde{\theta}\right\Vert ^{2}$
\cite[Thm. 1]{Patil.Fallin.ea2025}.
\end{lem}
Substituting (\ref{eq: Taylor approximation}) into (\ref{eq: CLES with f(x)})
yields
\begin{gather}
\dot{e}=\Phi^{\prime}\left(x,\hat{\theta}\right)\tilde{\theta}-k_{e}e-\frac{1}{2}\left(p+1\right)\gamma k_{T}\mu+\Delta\left(x,\tilde{\theta}\right),\label{eq: CLES with Taylor}
\end{gather}
where $\Delta:\RR^{n}\times\RR^{p}\to\RR^{n}$ is defined as $\text{\ensuremath{\Delta\left(x,\tilde{\theta}\right)}}\tq\varepsilon\left(x\right)+\mathcal{R}\left(x,\tilde{\theta}\right)$.
By using (\ref{thm: boundedness of desired trajectory}), (\ref{eq: tracking error}),
(\ref{eq: varepsilon bound}), Lemma \ref{lem: Lagrange remainder bound},
and the triangle inequality, the term $\Delta\left(x,\tilde{\theta}\right)$
can be bounded as 
\begin{equation}
\left\Vert \Delta\left(x,\tilde{\theta}\right)\right\Vert \leq\overline{\varepsilon}+\rho_{0}\left(\left\Vert e\right\Vert +\overline{x_{d}}\right)\left\Vert \tilde{\theta}\right\Vert ^{2}.\label{eq: Delta upperbound}
\end{equation}
Let $z:\RR_{\geq0}\to\RR^{\varphi}$ denote the concatenated error
defined as $z\tq\left[e^{\top},\,\tilde{\theta}^{\top}\right]^{\top},$
where $\varphi\tq n+p$. Since $\rho_{0}$ is strictly increasing,
the term $\rho_{0}\left(\left\Vert e\right\Vert +\overline{x_{d}}\right)$
can be upper-bounded as $\rho_{0}\left(\left\Vert e\right\Vert +\overline{x_{d}}\right)\leq\rho_{0}\left(\left\Vert z\right\Vert +\overline{x_{d}}\right)$.
Using this upper bound and concatenated error definition, (\ref{eq: Delta upperbound})
is further upper-bounded as
\begin{equation}
\left\Vert \Delta\left(x,\tilde{\theta}\right)\right\Vert \leq\overline{\varepsilon}+\rho_{1}\left(\left\Vert z\right\Vert \right)\left\Vert z\right\Vert ^{2},\label{eq: Final Delta upperbound}
\end{equation}
where $\rho_{1}\left(\left\Vert z\right\Vert \right)=\rho_{0}\left(\left\Vert z\right\Vert +\overline{x_{d}}\right)$.

Substituting (\ref{eq: CLES with f(x)}) into (\ref{eq: internal energy})
yields the closed-loop generalized internal energy{\small{}
\begin{equation}
U=e^{\top}\left(f\left(x\right)-k_{e}e-\Phi\left(x,\hat{\theta}\right)-\frac{1}{2}\left(p+1\right)\gamma k_{T}\mu\right)+\frac{1}{2}\sigma\hat{\theta}^{\top}\hat{\theta}.\label{eq: closed loop internal energy}
\end{equation}
}Substituting the gradient of (\ref{eq: closed loop internal energy})
in (\ref{eq: drift process - partial form}), yields the diffusion
term
\begin{gather}
\varrho=\Phi^{\prime\top}\left(x,\hat{\theta}\right)e+\frac{1}{2}\left(p+1\right)\gamma k_{T}e^{\top}\frac{\partial\mu}{\partial\hat{\theta}}-\sigma\hat{\theta}.\label{eq: Developed varrho}
\end{gather}
Substituting (\ref{eq: diffusion term}) and (\ref{eq: Developed varrho})
into (\ref{eq: Thermodynamic update law}) yields the LyLA-Therm update
law
\begin{align}
{\rm d}\hat{\theta} & =\gamma\text{proj}\bigg(\Phi^{\prime\top}\left(x,\hat{\theta}\right)e+\frac{1}{2}\left(p+1\right)\gamma k_{T}e^{\top}\frac{\partial\mu}{\partial\hat{\theta}}-\sigma\hat{\theta}\bigg){\rm d}t\nonumber \\
 & \hspace{1em}+\gamma\proj\left(\sqrt{k_{T}T\left(x,\hat{\theta},t\right)}{\rm d}\omega\right).\label{eq: Developed Lb-ThNN}
\end{align}
Taking the differential of the estimation error in (\ref{eq: Estimation error})
and using the definition of projection in (\ref{eq: projection definition})
on the diffusion term yields
\begin{align}
{\rm d}\tilde{\theta} & =-\gamma\text{proj}\bigg(\Phi^{\prime\top}\left(x,\hat{\theta}\right)e+\frac{1}{2}\left(p+1\right)\gamma k_{T}e^{\top}\frac{\partial\mu}{\partial\hat{\theta}}\nonumber \\
 & \hspace{1em}-\sigma\hat{\theta}\bigg){\rm d}t-\gamma\mathfrak{C}{\rm d}\omega,\label{eq: CLES estimation error - proj broke down}
\end{align}
where $\mathfrak{C}\in\RR_{\geq0}$ is defined as
\[
\mathfrak{C}\triangleq\begin{cases}
\sqrt{k_{T}T\left(x,\hat{\theta},t\right)}, & \text{Case 1},\\
\left(I-{\tt k}\left(\hat{\theta}\right)\frac{\nabla\mathcal{P}\nabla\mathcal{P}^{\top}}{\left\Vert \nabla\mathcal{P}\right\Vert ^{2}}\right)\sqrt{k_{T}T\left(x,\hat{\theta},t\right)}, & \text{Case 2},
\end{cases}
\]
where is $\hat{\theta}\in\overset{\circ}{\Pi}\text{ or }\nabla\mathcal{P}^{\top}\sqrt{k_{T}T\left(x,\hat{\theta},t\right)}{\rm d}\omega\leq0$
for Case 1, and $\hat{\theta}\in\Pi_{\varepsilon}\backslash\overset{\circ}{\Pi}\text{ and }\nabla\mathcal{P}^{\top}\sqrt{k_{T}T\left(x,\hat{\theta},t\right)}{\rm d}\omega>0$
for Case 2. 

Using (\ref{eq: tracking error}), (\ref{eq: Estimation error}),
and the chain rule, ${\rm d}z$ is obtained as ${\rm d}z=\left[\frac{{\rm d}e^{\top}}{{\rm d}t}{\rm d}t,\,{\rm d}\tilde{\theta}^{\top}\right]^{\top}$.
Substituting (\ref{eq: CLES with Taylor}) and (\ref{eq: CLES estimation error - proj broke down})
into ${\rm d}z$ yields
\begin{equation}
{\rm d}z=\begin{cases}
\mathcal{F}\left(z\right){\rm d}t+\mathcal{G}_{1}\left(z\right){\rm d}\omega, & \text{Case 1},\\
\mathcal{F}\left(z\right){\rm d}t+\mathcal{G}_{2}\left(z\right){\rm d}\omega, & \text{Case 2},
\end{cases}\label{eq: dz}
\end{equation}
where 
\begin{align*}
\mathcal{G}_{1}\left(z\right) & \tq\left[\begin{array}{c}
\boldsymbol{0}_{\left(\varphi-p\right)\times p}\\
-\gamma\sqrt{k_{T}T\left(x,\hat{\theta},t\right)}
\end{array}\right],\\
\mathcal{G}_{2}\left(z\right) & \tq\left[\begin{array}{c}
\boldsymbol{0}_{\left(\varphi-p\right)\times p}\\
-\gamma\big(I-{\tt k}\left(\hat{\theta}\right)\frac{\nabla\mathcal{P}\nabla\mathcal{P}^{\top}}{\left\Vert \nabla\mathcal{P}\right\Vert ^{2}}\big)\sqrt{k_{T}T\left(x,\hat{\theta},t\right)}
\end{array}\right].
\end{align*}
The term $\mathcal{F}$ in (\ref{eq: dz}) is defined as $\mathcal{F}\left(z\right)\tq\left[\mathcal{F}_{1}^{\top}\left(z\right),\mathcal{F}_{2}^{\top}\left(z\right)\right]^{\top}$,
where
\begin{align*}
\mathcal{F}_{1}\left(z\right) & \tq\Phi^{\prime}\left(x,\hat{\theta}\right)\tilde{\theta}-\frac{1}{2}\left(p+1\right)\gamma k_{T}\mu-k_{e}e+\Delta\left(x,\tilde{\theta}\right),\\
\mathcal{F}_{2}\left(z\right) & \tq\gamma\text{proj}\big(\begin{array}{c}
-\Phi^{\prime\top}\left(x,\hat{\theta}\right)e\end{array}\\
 & \hspace{1em}-\frac{1}{2}\left(p+1\right)\gamma k_{T}e^{\top}\frac{\partial\mu}{\partial\hat{\theta}}+\sigma\hat{\theta}\big).
\end{align*}

\section{Stability Analysis{\large{}\label{sec:Stability-Analysis}}}

Consider the Lyapunov function candidate $V_{L}:\mathcal{D}\to\RR_{\geq0}$
defined as 
\begin{equation}
V_{L}\left(z\right)\tq\frac{1}{2}e^{\top}e+\frac{1}{2\gamma}\tilde{\theta}^{\top}\tilde{\theta}.\label{eq: Lyapunov function candidate}
\end{equation}
The Lyapunov function candidate can be bounded as 
\begin{equation}
\alpha_{1}\left\Vert z\right\Vert ^{2}\leq V_{L}\left(z\right)\leq\alpha_{2}\left\Vert z\right\Vert ^{2},\label{eq: Raleigh Retz}
\end{equation}
where $\alpha_{1}\tq\frac{1}{2}\min\left(1,\frac{1}{\gamma}\right)$
and $\alpha_{2}\tq\frac{1}{2}\max\left(1,\frac{1}{\gamma}\right)$. 

For the universal function approximation property to hold, it is necessary
to ensure $x\in\Omega$, for all $t\in\RR_{\geq0}$. To guarantee
that $x\in\Omega$, it is subsequently shown that $z$ remains on
a compact domain. Consider the compact domain $\mathcal{D}\tq\left\{ \iota\in\RR^{\varphi}:\left\Vert \iota\right\Vert \leq\rho^{-1}\left({\tt b}_{0}-{\tt b}_{2}\right)\right\} $
in which $z$ is supposed to lie, where ${\tt b}_{0}\tq\min\left(\frac{k_{e}}{2}-\gamma{\tt c}_{2}\left(p+1\right)k_{T}\overline{\theta}-\frac{1}{2}\gamma\left({\tt c}_{1}\overline{\theta}+{\tt c_{0}}\right)\overline{\theta}\left(p+1\right)k_{T},\frac{\sigma}{2}\right)$,
${\tt b}_{2}\in\RR_{>0}$ denotes the desired rate of convergence,
${\tt c}_{0}$, ${\tt c}_{1}$, and ${\tt c}_{2}$ are defined in
(\ref{eq: bound on the partial of mu}), and $\rho\left(\left\Vert z\right\Vert \right)\tq\rho_{1}\left(\left\Vert z\right\Vert \right)\left\Vert z\right\Vert $
is invertible. If $\left\Vert z\right\Vert \in\mathcal{D}$, then
using (\ref{eq:vec_diff_prop}), (\ref{eq: tracking error}), and
the triangle inequality can be used to upper bound $x$ as
\begin{align*}
\left\Vert x\right\Vert  & \leq\left\Vert e\right\Vert +\overline{x_{d}}\leq\left\Vert z\right\Vert +\overline{x_{d}}\leq\rho^{-1}\left({\tt b}_{0}-{\tt b}_{2}\right)+\overline{x_{d}}.
\end{align*}
Then $z\in\mathcal{D}$ implies $x\in\Omega$, where $\Omega$ is
selected as $\Omega\triangleq\left\{ \iota\in\RR^{n}:\left\Vert \iota\right\Vert \leq\rho^{-1}\left({\tt b}_{0}-{\tt b}_{2}\right)+\overline{x_{d}}\right\} $.
In the subsequent analysis, it is shown that if $z\left(0\right)\in\mathcal{S}$,
then $z\left(t\right)\in\mathcal{D}$, for all $t\in\left[0,\infty\right)$,
where $\mathcal{S}\subset\mathcal{D}$ is defined as
\begin{equation}
\mathcal{S}\tq\left\{ \iota\in\RR^{\varphi}:\left\Vert \iota\right\Vert \leq\sqrt{\frac{\alpha_{1}}{\alpha_{2}}\left(\rho^{-1}\left({\tt b}_{0}-{\tt b}_{2}\right)\right)^{2}-\frac{{\tt b}_{1}}{{\tt b}_{2}}}\right\} ,\label{eq:Chi Definition}
\end{equation}
where ${\tt b}_{1}\tq\frac{\overline{\varepsilon}^{2}}{2k_{e}}+\frac{1}{2}\gamma\left({\tt c}_{1}\overline{\theta}+{\tt c_{0}}\right)\overline{\theta}\left(p+1\right)k_{T}+\frac{1}{2}\sigma\overline{\theta}^{2}$.
Note that the set $\mathcal{S}$ exists when ${\tt b}_{0}>{\tt b}_{2}+\rho\left(\sqrt{\frac{\alpha_{2}}{\alpha_{1}}\frac{{\tt b}_{1}}{{\tt b}_{2}}}\right)$.
Additionally, $\mathcal{S}$ can be made arbitrarily large to include
any initial condition $z\left(0\right)$ such that ${\tt b}_{0}$
satisfies the gain condition
\begin{equation}
{\tt b}_{0}\geq\rho\left(\sqrt{\frac{\alpha_{2}}{\alpha_{1}}\left\Vert z\right\Vert ^{2}+\frac{\alpha_{2}}{\alpha_{1}}\frac{{\tt b}_{1}}{{\tt b}_{2}}}\right)+{\tt b}_{2}.\label{eq: gain condition}
\end{equation}
Note that the gain condition in (\ref{eq: gain condition}) is equivalent
to stating the set $\mathcal{S}$ is not a null set and contains the
initial condition $z\left(0\right)$.

Since $\mathcal{S}\subset\mathcal{D}$, $z\left(0\right)\in\mathcal{S}$
implies $z\left(0\right)\in\mathring{\mathcal{D}}$ where the notation
$\mathring{\mathcal{D}}$ denotes the interior of $\mathcal{D}$.
The solution $t\mapsto z\left(t\right)$ is assumed to be continuous
on a time-interval $\mathcal{I}\triangleq\left[0,t_{1}\right]$ with
probability one\footnote{This is a standard assumption in the analysis of stochastic systems
(cf. \cite[Page 36, Ass. (A2)]{Kushner1971})} such that $z\left(t\right)\in\mathcal{D}$ for all $t\in\mathcal{I}$.
It follows that $x\left(t\right)\in\Omega$, for all $t\in\mathcal{I}$,
therefore the universal function approximation property holds over
this time interval. In the subsequent stability analysis, the probabilistic
convergence properties of the solutions are analyzed. To facilitate
the analysis, the risk of $z$ to escape $\mathcal{D}$ is denoted
as $\vartheta$, which is defined as
\begin{align}
\vartheta & \tq\frac{1}{\alpha_{1}\left(\rho^{-1}\left({\tt b}_{0}-{\tt b}_{2}\right)\right)^{2}}V_{L}\left(z\left(0\right)\right)\nonumber \\
 & \quad+\frac{1}{\lambda}V_{L}\left(z\left(0\right)\right)\exp\left(-{\tt b}_{1}t\right)+\frac{\alpha_{2}{\tt b}_{1}}{\lambda{\tt b}_{2}},\label{eq: Escape risk}
\end{align}
 where $\lambda\in\left[\frac{\alpha_{2}{\tt b}_{1}}{{\tt b}_{2}},\alpha_{1}\left(\rho^{-1}\left({\tt b}_{0}-{\tt b}_{2}\right)\right)^{2}\right].$
\begin{thm}
\label{thm: Stability }Consider the dynamical system in (\ref{eq: dynamical system}).
For any $z\left(0\right)\in\mathcal{S}$, such that (\ref{eq: gain condition})
holds, the update law and the controller given by (\ref{eq: Thermodynamic update law})
and (\ref{eq: controller}) ensures that the solution $z\left(t\right)$
is uniformly ultimately bounded in probability (UUB-p, \cite[Def. 1]{Akbari.Nino.ea2024})
in the sense that
\begin{equation}
{\rm P}\left(\underset{t\leq s<\infty}{\sup}\left\Vert z\left(s\right)\right\Vert \leq\sqrt{\frac{\lambda}{\alpha_{1}}}\right)\geq1-\vartheta.\label{eq: Stability result}
\end{equation}
\end{thm}
\begin{IEEEproof}
Since the closed-loop error system in (\ref{eq: dz}) consists of
two cases with different diffusion matrices, the proof is divided
into two cases. Case 1 where ${\tt k}\left(\theta\right)<0,\text{ or }\nabla{\tt k}\left(\theta\right)^{\top}\sqrt{k_{T}T\left(x,\hat{\theta},t\right)}{\rm d}\omega\leq0$,
and Case 2 where ${\tt k}\left(\theta\right)=0,\text{ and }\nabla{\tt k}\left(\theta\right)^{\top}\sqrt{k_{T}T\left(x,\hat{\theta},t\right)}{\rm d}\omega>0$.

\uline{Case 1}: Taking the infinitesimal generator of the candidate
Lyapunov function in (\ref{eq: Lyapunov function candidate}) for
Case 1 of (\ref{eq: dz}) yields (cf., \cite[Eq. 7]{Akbari.Nino.ea2024})
\begin{equation}
\mathcal{L}V_{L}\left(z\right)=\frac{\partial V_{L}}{\partial z}\mathcal{F}\left(z\right)+\frac{1}{2}\tr\left(\mathcal{G}_{1}^{\top}\left(z\right)\frac{\partial^{2}V_{L}}{\partial z^{2}}\mathcal{G}_{1}\left(z\right)\right).\label{eq: General fomulation of L differential}
\end{equation}
Substituting $\mathcal{F}\left(z\right)$ and $\mathcal{G}_{1}\left(z\right)$
into (\ref{eq: General fomulation of L differential}) yields
\begin{align}
\mathcal{L}V_{L}\left(z\right) & =e^{\top}\bigg(\Phi^{\prime}\left(x,\hat{\theta}\right)\tilde{\theta}-\frac{1}{2}\left(p+1\right)\gamma k_{T}\mu-k_{e}e\nonumber \\
 & \hspace{1em}+\Delta\left(x,\tilde{\theta}\right)\bigg)-\tilde{\theta}^{\top}\text{proj}\bigg(\Phi^{\prime\top}\left(x,\hat{\theta}\right)e\nonumber \\
 & \hspace{1em}+\frac{1}{2}\left(p+1\right)\gamma k_{T}e^{\top}\frac{\partial\mu}{\partial\hat{\theta}}-\sigma\hat{\theta}\bigg)\nonumber \\
 & \hspace{1em}+\frac{1}{2\gamma}\text{tr}\left(\gamma^{2}\sqrt{k_{T}T\left(x,\hat{\theta},t\right)}^{\top}\sqrt{k_{T}T\left(x,\hat{\theta},t\right)}\right).\label{eq: Substituted F and G into LV}
\end{align}
From \cite[P2 in Thm. 1]{Cai2006a}, $\left(\cdot\right)\leq\text{proj}\left(\cdot\right)$.
Using this lower bound of proj$\left(\cdot\right)$, (\ref{eq: Temperature law}),
and the fact that $\sqrt{k_{T}T\left(x,\hat{\theta},t\right)}$ is
a scalar and $T$ is designed to be positive, (\ref{eq: Substituted F and G into LV})
is simplified as
\begin{align}
\mathcal{L}V_{L}\left(z\right) & \leq e^{\top}\left(-k_{e}e+\Delta\left(x,\tilde{\theta}\right)\right)\nonumber \\
 & \hspace{1em}-\tilde{\theta}^{\top}\left(\frac{1}{2}\left(p+1\right)\gamma k_{T}e^{\top}\frac{\partial\mu}{\partial\hat{\theta}}-\sigma\hat{\theta}\right).\label{eq: Substituted det and canceling}
\end{align}

\uline{Case 2}: Taking the infinitesimal generator of the candidate
Lyapunov function in (\ref{eq: Lyapunov function candidate}) with
respect to Case 2 of (\ref{eq: dz}) yields (cf., \cite[Eq. 7]{Akbari.Nino.ea2024})
\begin{equation}
\mathcal{L}V_{L}\left(z\right)=\frac{\partial V_{L}}{\partial z}\mathcal{F}\left(z\right)+\frac{1}{2}\tr\left(\mathcal{G}_{2}^{\top}\left(z\right)\frac{\partial^{2}V_{L}}{\partial z^{2}}\mathcal{G}_{2}\left(z\right)\right).\label{eq: General fomulation of L differential-1}
\end{equation}
Substituting $\mathcal{F}\left(z\right)$ and $\mathcal{G}_{2}\left(z\right)$
into (\ref{eq: General fomulation of L differential}) yields
\begin{align}
\mathcal{L}V_{L}\left(z\right) & =e^{\top}\bigg(\Phi^{\prime}\left(x,\hat{\theta}\right)\tilde{\theta}-\frac{1}{2}\left(p+1\right)\gamma k_{T}\mu-k_{e}e\nonumber \\
 & +\Delta\left(x,\tilde{\theta}\right)\bigg)-\tilde{\theta}^{\top}\text{proj}\bigg(\Phi^{\prime\top}\left(x,\hat{\theta}\right)e\nonumber \\
 & +\frac{1}{2}\left(p+1\right)\gamma k_{T}e^{\top}\frac{\partial\mu}{\partial\hat{\theta}}-\sigma\hat{\theta}\bigg)\nonumber \\
 & +\frac{1}{2\gamma}\tr\bigg(\gamma^{2}\big(I-\frac{{\tt k}\left(\hat{\theta}\right)\nabla\mathcal{P}\nabla\mathcal{P}^{\top}}{\left\Vert \nabla\mathcal{P}\right\Vert ^{2}}\big)^{\top}\sqrt{k_{T}T\left(x,\hat{\theta},t\right)}\nonumber \\
 & \left(I-{\tt k}\left(\hat{\theta}\right)\frac{\nabla\mathcal{P}\nabla\mathcal{P}^{\top}}{\left\Vert \nabla\mathcal{P}\right\Vert ^{2}}\right)\sqrt{k_{T}T\left(x,\hat{\theta},t\right)}\bigg).\label{eq: Substituted F and G into LV-1}
\end{align}
Again using the projection lower bound, applying (\ref{eq: order of multiplication property})
to $\tr\left({\tt k}^{2}\left(\hat{\theta}\right)k_{T}T\left(x,\hat{\theta},t\right)\frac{\nabla\mathcal{P}\nabla\mathcal{P}^{\top}}{\left\Vert \nabla\mathcal{P}\right\Vert ^{2}}\right),$
and canceling the cross terms yields 
\begin{align}
\mathcal{L}V_{L}\left(z\right) & \leq e^{\top}\left(-\frac{1}{2}\left(p+1\right)\gamma k_{T}\mu-k_{e}e+\Delta\left(x,\tilde{\theta}\right)\right)\nonumber \\
 & -\tilde{\theta}^{\top}\left(\frac{1}{2}\left(p+1\right)\gamma k_{T}e^{\top}\frac{\partial\mu}{\partial\hat{\theta}}-\sigma\hat{\theta}\right)\nonumber \\
 & +\frac{1}{2}\gamma\text{tr}\left(k_{T}T\left(x,\hat{\theta},t\right)I_{p}\right)\nonumber \\
 & +\frac{1}{2}\gamma\tr\left({\tt k}^{2}\left(\hat{\theta}\right)k_{T}T\left(x,\hat{\theta},t\right)\right)\nonumber \\
 & -\frac{1}{2}\gamma\tr\left(2{\tt k}\left(\hat{\theta}\right)k_{T}T\left(x,\hat{\theta},t\right)\frac{\nabla\mathcal{P}\nabla\mathcal{P}^{\top}}{\left\Vert \nabla\mathcal{P}\right\Vert ^{2}}\right).\label{eq: LV_L case 2 - first step simplified}
\end{align}
Applying the facts that $\tr\left(2{\tt k}\left(\hat{\theta}\right)k_{T}T\left(x,\hat{\theta},t\right)\frac{\nabla\mathcal{P}\nabla\mathcal{P}^{\top}}{\left\Vert \nabla\mathcal{P}\right\Vert ^{2}}\right)$
is positive and ${\tt k}^{2}\left(\hat{\theta}\right)\leq1$ to (\ref{eq: LV_L case 2 - first step simplified})
yields
\begin{align}
\mathcal{L}V_{L}\left(z\right) & \leq e^{\top}\left(-\frac{1}{2}\left(p+1\right)\gamma k_{T}\mu-k_{e}e+\Delta\left(x,\tilde{\theta}\right)\right)\nonumber \\
 & -\tilde{\theta}^{\top}\left(\frac{1}{2}\left(p+1\right)\gamma k_{T}e^{\top}\frac{\partial\mu}{\partial\hat{\theta}}-\sigma\hat{\theta}\right)\nonumber \\
 & +\frac{1}{2}\gamma\left(p+1\right)k_{T}T\left(x,\hat{\theta},t\right)+\frac{1}{2}\gamma k_{T}T\left(x,\hat{\theta},t\right).\label{eq: LV_L case 2 - second step simplified}
\end{align}
Applying (\ref{eq: Temperature law}) to (\ref{eq: LV_L case 2 - second step simplified})
and then canceling the cross terms yields (\ref{eq: Substituted det and canceling}).
Therefore, both Cases 1 and 2 yield (\ref{eq: Substituted det and canceling}). 

Using (\ref{eq: bound on the partial of mu}), (\ref{eq: Estimation error}),
and the bound on $\Delta\left(x,\tilde{\theta}\right)$ in (\ref{eq: Final Delta upperbound})
on (\ref{eq: Substituted det and canceling}) yields
\begin{align}
\mathcal{L}V_{L}\left(z\right) & \leq-k_{e}\left\Vert e\right\Vert ^{2}+\overline{\varepsilon}\left\Vert e\right\Vert +\rho_{1}\left(\left\Vert z\right\Vert \right)\left\Vert z\right\Vert ^{2}\left\Vert e\right\Vert \nonumber \\
 & +\frac{1}{2}\gamma{\tt c}_{2}\left(p+1\right)k_{T}\left\Vert \tilde{\theta}\right\Vert \left\Vert e\right\Vert ^{2}\nonumber \\
 & +\frac{1}{2}\gamma{\tt c}_{1}\left(p+1\right)k_{T}\left\Vert \tilde{\theta}\right\Vert \left\Vert e\right\Vert \left\Vert \hat{\theta}\right\Vert \nonumber \\
 & +\frac{1}{2}\gamma{\tt c}_{0}\left(p+1\right)k_{T}\left\Vert \tilde{\theta}\right\Vert \left\Vert e\right\Vert -\sigma\left\Vert \tilde{\theta}\right\Vert ^{2}+\sigma\left\Vert \tilde{\theta}\right\Vert \left\Vert \theta^{*}\right\Vert ,\label{eq: LV_L after bound of higher order}
\end{align}
for all $t\in\mathcal{I}$. Using the bound of $\theta^{*}$ in Section
\ref{subsec:Thermodynamic-Neural-Network}, the bound of $\hat{\theta}$
in Lemma \ref{lem: boundedness of theta hat}, and (\ref{eq: Estimation error}),
$\mathcal{L}V_{L}$ in (\ref{eq: LV_L after bound of higher order})
is further upper bounded as
\begin{align}
\mathcal{L}V_{L}\left(z\right) & \leq-\left(k_{e}-\gamma{\tt c}_{2}\left(p+1\right)k_{T}\overline{\theta}\right)\left\Vert e\right\Vert ^{2}\nonumber \\
 & +\overline{\varepsilon}\left\Vert e\right\Vert +\rho_{1}\left(\left\Vert z\right\Vert \right)\left\Vert z\right\Vert ^{3}\nonumber \\
 & +\gamma\left({\tt c}_{1}\overline{\theta}+{\tt c_{0}}\right)\overline{\theta}\left(p+1\right)k_{T}\left\Vert e\right\Vert \nonumber \\
 & -\sigma\left\Vert \tilde{\theta}\right\Vert ^{2}+\sigma\overline{\theta}\left\Vert \tilde{\theta}\right\Vert ,\label{eq: LV_L after bound of higher order-1}
\end{align}
for all $t\in\mathcal{I}$. Using Young's inequality on $\overline{\varepsilon}\left\Vert e\right\Vert $,
$\sigma\overline{\theta}\left\Vert \tilde{\theta}\right\Vert $, and
$\gamma\left({\tt c}_{1}\overline{\theta}+{\tt c_{0}}\right)\overline{\theta}\left(p+1\right)k_{T}\left\Vert e\right\Vert $
yields
\begin{equation}
\mathcal{L}V_{L}\leq-\left({\tt b}_{0}-\rho\left(\left\Vert z\right\Vert \right)\right)\left\Vert z\right\Vert ^{2}+{\tt b}_{1},\label{eq: LV_L almost final}
\end{equation}
for all $t\in\mathcal{I}$. Recall that $z\left(t\right)\in\mathcal{D},$
for all $t\in\mathcal{I}$. Therefore, from the definition of $\mathcal{D}$,
it follows that $\rho\left(\left\Vert z\right\Vert \right)\leq{\tt b}_{0}-{\tt b}_{2}$
for all $t\in\mathcal{I}$. Therefore,
\begin{equation}
\mathcal{L}V_{L}\leq-{\tt b}_{2}\left\Vert z\right\Vert ^{2}+{\tt b}_{1},\label{eq: LV_L are we there yet}
\end{equation}
for all $t\in\mathcal{I}$. Applying (\ref{eq: Raleigh Retz}) to
(\ref{eq: LV_L are we there yet}) yields
\begin{equation}
\mathcal{L}V_{L}\leq-\frac{{\tt b}_{2}}{\alpha_{2}}V_{L}+{\tt b}_{1},\label{eq: LV_L FINALLL!!-1}
\end{equation}
for all $t\in\mathcal{I}$.

Since $V_{L}\left(0\right)=0$, $V_{L}\in\mathcal{C}^{2}$, and $z$
is a continuous strong Markov process, based on (\ref{eq: LV_L FINALLL!!-1}),
then \cite[Lemma 1]{Akbari.Nino.ea2024} can be invoked to state 
\[
{\rm P}\left(\underset{t\leq s\leq\infty}{\sup}V_{L}\left(z\left(s\right)\right)\geq\lambda\right)\leq\vartheta,
\]
which is equivalent to 
\begin{equation}
{\rm P}\left(\underset{t\leq s\leq\infty}{\sup}V_{L}\left(z\left(s\right)\right)<\lambda\right)\leq1-\vartheta,\label{eq: probability of VL being bounded}
\end{equation}
for all $t\in\mathcal{I}$. From (\ref{eq: Raleigh Retz}), ${\rm P}\left(\underset{t\leq s<\infty}{\sup}\left\Vert z\left(s\right)\right\Vert ^{2}<\frac{\lambda}{\alpha_{1}}\right)\geq{\rm P}\left(\underset{t\leq s<\infty}{\sup}V_{L}\left(z\left(s\right)\right)<\lambda\right)$.
Therefore, using (\ref{eq: probability of VL being bounded}) and
Lemma \ref{thm:probability} yields (\ref{eq: Stability result}),
for all $t\in\mathcal{I}$. From (\ref{eq: Stability result}) and
\cite[Def. 1]{Akbari.Nino.ea2024}, the solution $z\left(t\right)$
is UUB-p, for all $t\in\mathcal{I}$.

Let ${\tt S}_{1}\tq\left\{ z:\left\Vert z\right\Vert <\sqrt{\frac{\lambda}{\alpha_{1}}}\right\} $,
and ${\tt S}_{2}\tq\left\{ z:\left\Vert e\right\Vert <\sqrt{\frac{\lambda}{\alpha_{1}}}\right\} $.
Since ${\tt S}_{1}\subseteq{\tt S}_{2}$, the monotonicity property
in (\ref{eq:monotonicity}) yields ${\rm P}\left(\underset{t\leq s<\infty}{\sup}\left\Vert z\right\Vert <\sqrt{\frac{\lambda}{\alpha_{1}}}\right)\leq{\rm P}\left(\underset{t\leq s<\infty}{\sup}\left\Vert e\right\Vert <\sqrt{\frac{\lambda}{\alpha_{1}}}\right)$.
This inequality together with (\ref{eq: Stability result}) yields
${\rm P}\left(\underset{t\leq s<\infty}{\sup}\left\Vert e\right\Vert <\sqrt{\frac{\lambda}{\alpha_{1}}}\right)\geq1-\vartheta$.
Let ${\tt S}_{3}\tq\left\{ z:\left\Vert x\right\Vert <\sqrt{\frac{\lambda}{\alpha_{1}}}+\overline{x_{d}}\right\} $.
Since ${\tt S}_{2}\subseteq{\tt S}_{3}$, the monotonicity property
in (\ref{eq:monotonicity}) yields ${\rm P}\left(\underset{t\leq s<\infty}{\sup}\left\Vert x\left(s\right)\right\Vert <\sqrt{\frac{\lambda}{\alpha_{1}}}+\overline{x_{d}}\right)\geq{\rm P}\left(\underset{t\leq s<\infty}{\sup}\left\Vert e\left(s\right)\right\Vert <\sqrt{\frac{\lambda}{\alpha_{1}}}\right)\geq1-\vartheta$.
Let ${\tt S}_{4}\tq\left\{ z:\left\Vert \tilde{\theta}\right\Vert <\sqrt{\frac{\lambda}{\alpha_{1}}}\right\} $.
Since ${\tt S}_{1}\subseteq{\tt S}_{4}$, the monotonicity property
in (\ref{eq:monotonicity}) yields ${\rm P}\left(\underset{t\leq s<\infty}{\sup}\left\Vert \tilde{\theta}\left(s\right)\right\Vert <\sqrt{\frac{\lambda}{\alpha_{1}}}\right)\geq{\rm P}\left(\underset{t\leq s<\infty}{\sup}\left\Vert z\left(s\right)\right\Vert <\sqrt{\frac{\lambda}{\alpha_{1}}}\right)$.
This obtained inequality together with (\ref{eq: Stability result})
yields ${\rm P}\left(\underset{t\leq s<\infty}{\sup}\left\Vert \tilde{\theta}\left(s\right)\right\Vert <\sqrt{\frac{\lambda}{\alpha_{1}}}\right)\geq1-\vartheta$.
Since ${\rm P}\left\{ \underset{t\leq s<\infty}{\sup}\left\Vert x\left(s\right)\right\Vert <\sqrt{\frac{\lambda}{\alpha_{1}}}+\overline{x_{d}}\right\} \geq1-\vartheta$
and $\hat{\theta}$ is bounded, and based on the smoothness of $\Phi\left(x,\hat{\theta}\right)$,
there exists a constant $\overline{\Phi}\in\RR_{>0}$ such that ${\rm P}\left(\underset{t\leq s<\infty}{\sup}\left\Vert \Phi\left(x\left(s\right),\hat{\theta}\left(s\right)\right)\right\Vert \leq\overline{\Phi}\right)\geq1-\vartheta$.
Since ${\rm P}\left(\underset{t\leq s<\infty}{\sup}\left\Vert \Phi\left(x\left(s\right),\hat{\theta}\left(s\right)\right)\right\Vert \leq\overline{\Phi}\right)\geq1-\vartheta$,
${\rm P}\left(\underset{t\leq s<\infty}{\sup}\left\Vert e\left(s\right)\right\Vert <\sqrt{\frac{\lambda}{\alpha_{1}}}\right)\geq1-\vartheta$,
using Assumption \ref{thm: boundedness of desired trajectory}, (\ref{eq: controller}),
and boundedness of $\mu$ yields ${\rm P}\left(\underset{t\leq s<\infty}{\sup}\left\Vert u\left(s\right)\right\Vert \leq\overline{u}\right)\geq1-\vartheta$,
for some constant $\overline{u}\in\RR_{>0}$. Therefore, all implemented
signals are bounded with probability of $1-\vartheta$.
\end{IEEEproof}

\section{Simulation}

To assess the efficacy of the proposed LyLA-Therm adaptive controller,
simulations are conducted on a five-dimensional nonlinear dynamical
system, where the function $f$ in (\ref{eq: dynamical system}) is
defined as
\[
f=\left[\begin{array}{c}
5\tanh\left(50x_{1}\right)x_{5}^{2}+\cos\left(x_{4}\right)\\
\cos\left(20x_{3}\right)+2\sin\left(x_{1}x_{2}\right)\sin\left(x_{4}x_{5}\right)\\
10\exp\left(-25x_{4}^{2}\right)x_{3}-0.1x_{3}^{3}\\
2\sin\left(15\left(x_{1}x_{5}-x_{2}x_{3}\right)\right)\\
-x_{1}x_{5}+5\tanh\left(20\left(x_{2}-x_{4}\right)\right)
\end{array}\right],
\]
and $x\tq\left[x_{1},x_{2},x_{3},x_{4},x_{5}\right]^{\top}:\RR_{\geq0}\to\RR^{5}$
denotes the system state, and $g=I_{5\times5}$. The simulations are
performed over $30$ seconds and initial condition $x\left(0\right)=\left[0,-1,3,-3,3\right]^{\top}$.
The desired trajectory is selected as $x_{d}=[\sin\left(2t\right),-\cos\left(t\right),\sin\left(3t\right)+\cos\left(-2t\right),\sin\left(t\right)-\cos\left(-0.5t\right),\sin\left(-t\right)].$
Four simulations are performed. The first simulation (S1) represents
the performance of the baseline Lb-DNN similar to the one used in
\cite{Akbari.Nino.ea2024}, which employs a deterministic update law.
The second (S2), third (S3), and fourth (S4) simulations implement
LyLA-Therm architectures with three different values of $\mu$ and
consequently different generalized temperature laws and diffusion
terms. The parameter $\mu$ for the second, third, and fourth simulations
is denoted as $\mu_{2}$, $\mu_{3}$, and $\mu_{4}$ respectively,
and defined as $\mu_{2}\triangleq9e,$ $\mu_{3}\triangleq e\left(0.01\left\Vert x\right\Vert ^{2}+9\right),$
and $\mu_{4}\triangleq e\left(0.01\left\Vert \hat{\theta}\right\Vert ^{2}+9\right).$
For S2-S4, the Wiener process, $\omega$, is generated with mean of
0 and covariance of 1. For all four simulations the learning rates
and forgetting factors are selected as $\gamma=1$ and $\sigma=0.001$,
respectively. The gain $k_{T}$ introduced in (\ref{eq: general form of varsigma})
is selected as $k_{T}=0.03$, and the control gain in (\ref{eq: controller})
is selected as $k_{e}=100$. For all simulations, the deep learning
architectures have $k=9$ inner layers with $L=10$ neurons per hidden
layer, and the architectures use swish activation functions (see \cite{Ramachandran.Zoph.ea2017}).
Since Swish activation, a smooth approximation of ReLU activation,
is used for the simulations, the weight estimates are initialized
via Kaiming He initialization (see \cite{He.Zhang.ea2015}), which
is specifically designed as the standard initialization method for
ReLU-based networks and remains appropriate here since Swish closely
approximates ReLU behavior.

The simulations compare the performance of tracking error, function
approximation error, and off-trajectory function approximation errors
of the baseline Lb-DNN used in \cite{Akbari.Nino.ea2024} against
different cases of LyLA-Therm architectures. For the off-trajectory
function approximation comparison, the simulations examine the RMS
function approximation error performance on a test dataset comprising
$90$ random off-trajectory points selected from the uniform distribution
$U\left(-0.5,0.5\right)$ to evaluate the function approximation capability
of the architecture beyond the explored points.

\begin{table}
\begin{centering}
\par\end{centering}
\begin{centering}
\caption{\label{tab: Performance comparison results}Performance comparison
results}
\par\end{centering}
\centering{}%
\begin{tabular}{c|c|c|c}
\multicolumn{1}{c}{} & \multirow{2}{*}{RMS $\left\Vert e\right\Vert $} & \multirow{2}{*}{RMS $\left\Vert f-\Phi\left(x,\hat{\theta}\right)\right\Vert $} & Off-trajectory RMS\tabularnewline
\multicolumn{1}{c}{} &  &  & $\left\Vert f-\Phi\left(x,\hat{\theta}\right)\right\Vert $\tabularnewline
\hline 
S1 & $0.0734$ & $7.2575$ & $5.7432$\tabularnewline
\hline 
S2 & $0.0586$ & $5.7656$ & $5.0960$\tabularnewline
\hline 
S3 & $0.0587$ & $5.7743$ & $5.0937$\tabularnewline
\hline 
S4 & $0.0582$ & $5.7412$ & $5.3619$\tabularnewline
\end{tabular}
\end{table}

Based on Table \ref{tab: Performance comparison results}, the tracking
errors for S2-S4 demonstrate improvements of 20.0875\%, 19.9662\%,
and 20.6629\% over the baseline Lb-DNN (S1), respectively. The function
approximation errors for S2-S4 show improvements of 20.5572\%, 20.4375\%,
and 20.8936\% over the baseline Lb-DNN (S1), respectively. The off-trajectory
function approximation errors for S2-S4 exhibit improvements of 11.2688\%,
11.3087\%, and 6.6402\% over the baseline Lb-DNN (S1), respectively.

\section{Conclusion}

This paper introduced a novel Lyapunov-based Langevin Adaptive Thermodynamic
neural network controller that leverages thermodynamic principles
to balance exploration and exploitation in parameter adaptation. By
formulating the update law as a Langevin-type stochastic differential
equation, the developed approach ensures controlled stochasticity,
where the drift term minimizes generalized internal energy for exploitation
and the diffusion term, governed by a user-designed generalized temperature
law, facilitates exploration. Additionally, an adaptive controller
is designed that assists the exploitation objective as well as compensating
for the effect of the stochasticity. Through a stochastic Lyapunov-based
analysis, the tracking error and parameter estimation errors are guaranteed
to exponentially converge to an ultimate bound in probability, while
allowing for flexibility of design in the generalized temperature
law.

\bibliographystyle{ieeetr}
\bibliography{ncr,master,encr}

\begin{IEEEbiography}[{\includegraphics[scale=0.09]{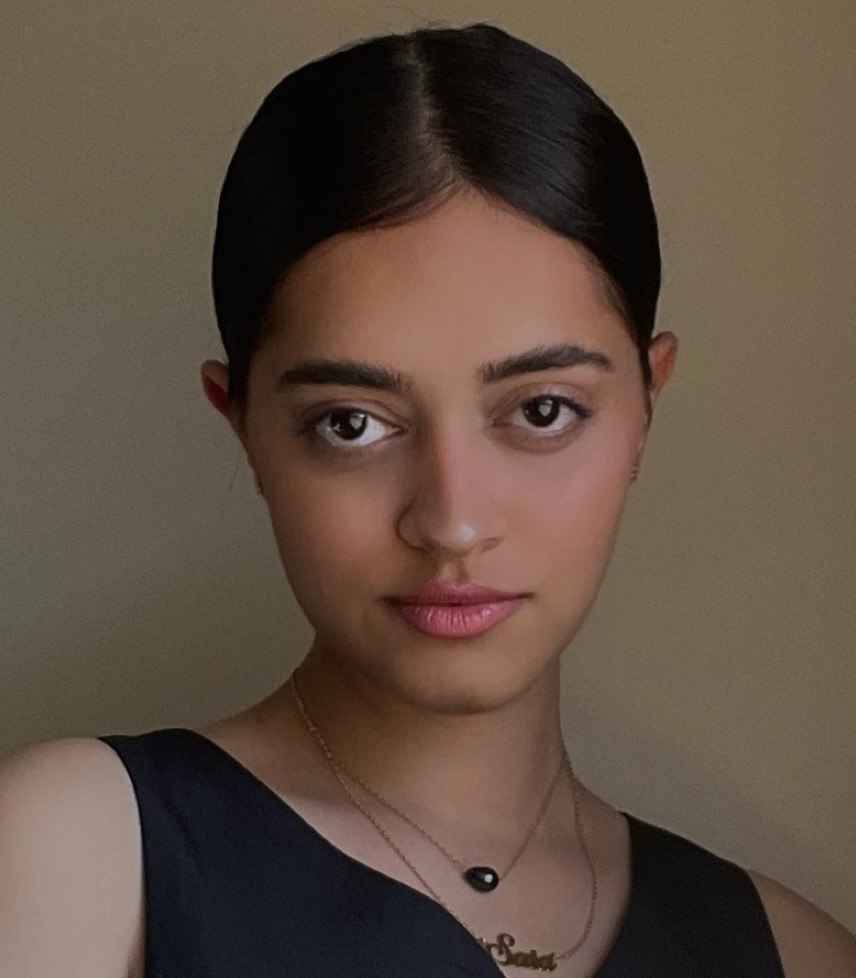}}]{Saiedeh Akbari}
 is a Ph.D. candidate in the department of Mechanical and Aerospace
Engineering at University of Florida. Her research focuses on developing
adaptive learning-based control strategies for stochastic nonlinear
systems. Saiedeh completed her Bachelor of Science in Mechanical Engineering
at KN Toosi University of Technology in 2020. During her undergraduate
studies, she conducted research on discrete-time sliding mode control
for permanent magnet DC motors. She then received her Master of Science
in Mechanical Engineering at The University of Alabama in 2023, where
she worked on nonlinear adaptive control techniques for hybrid systems,
with applications in rehabilitation robotics.
\end{IEEEbiography}

\begin{IEEEbiography}[{\includegraphics[scale=0.105]{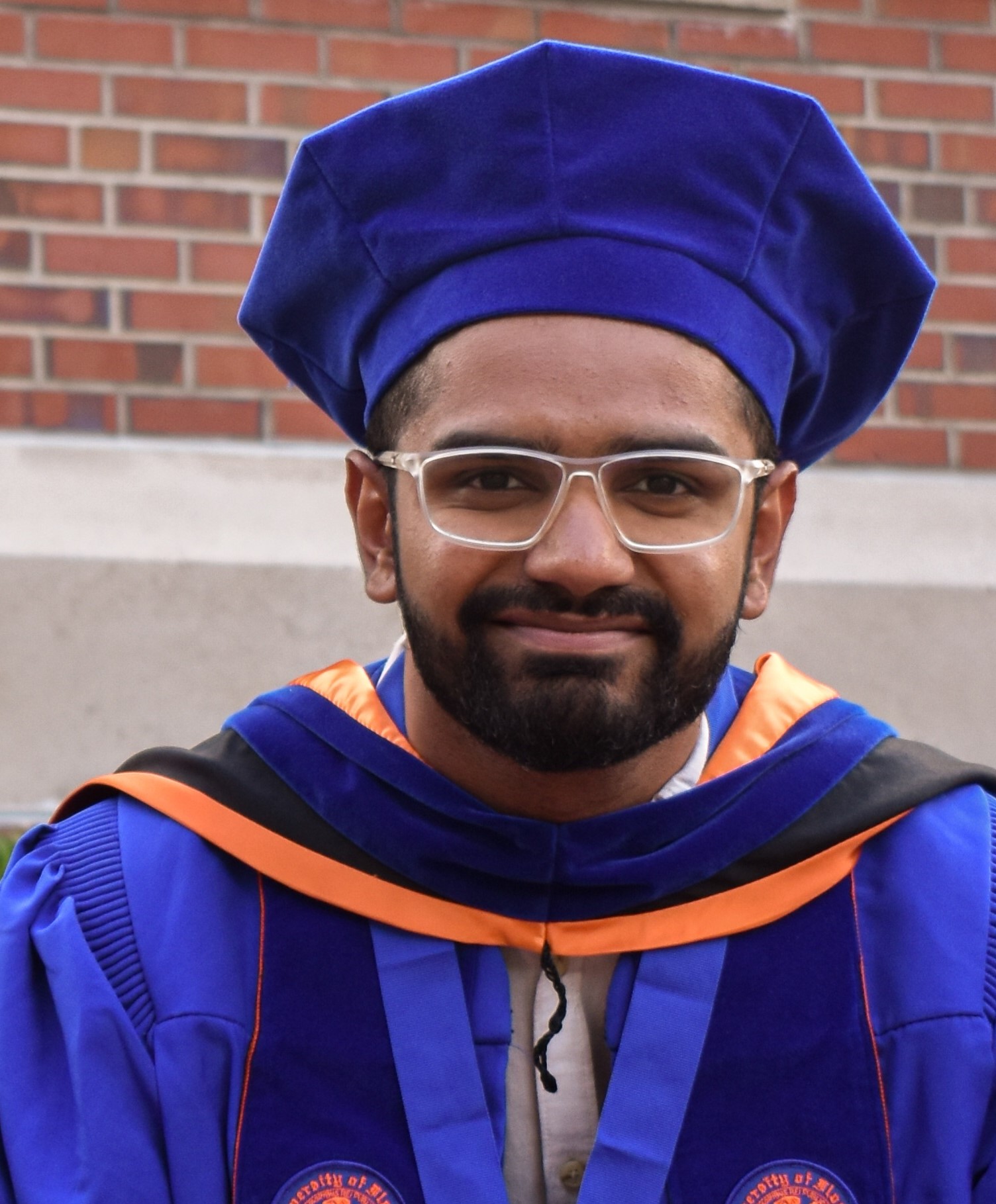}}]{Omkar Sudhir Patil}
 received his Bachelor of Technology (B.Tech.) degree in production
and industrial engineering from Indian Institute of Technology (IIT)
Delhi in 2018, where he was honored with the BOSS award for his outstanding
bachelor's thesis project. In 2019, he joined the Nonlinear Controls
and Robotics (NCR) Laboratory at the University of Florida under the
guidance of Dr. Warren Dixon to pursue his doctoral studies. Omkar
received his Master of Science (M.S.) degree in mechanical engineering
in 2022 and Ph.D. in mechanical engineering in 2023 from the University
of Florida. During his PhD studies, he was awarded the Graduate Student
Research Award for outstanding research. In 2023, he started working
as a postdoctoral research associate at NCR Laboratory, University
of Florida. His research focuses on the development and application
of innovative Lyapunov-based nonlinear, robust, and adaptive control
techniques. 
\end{IEEEbiography}

\begin{IEEEbiography}[{\includegraphics[viewport=131.291bp 0bp 1425.445bp 1487bp,clip,scale=0.061]{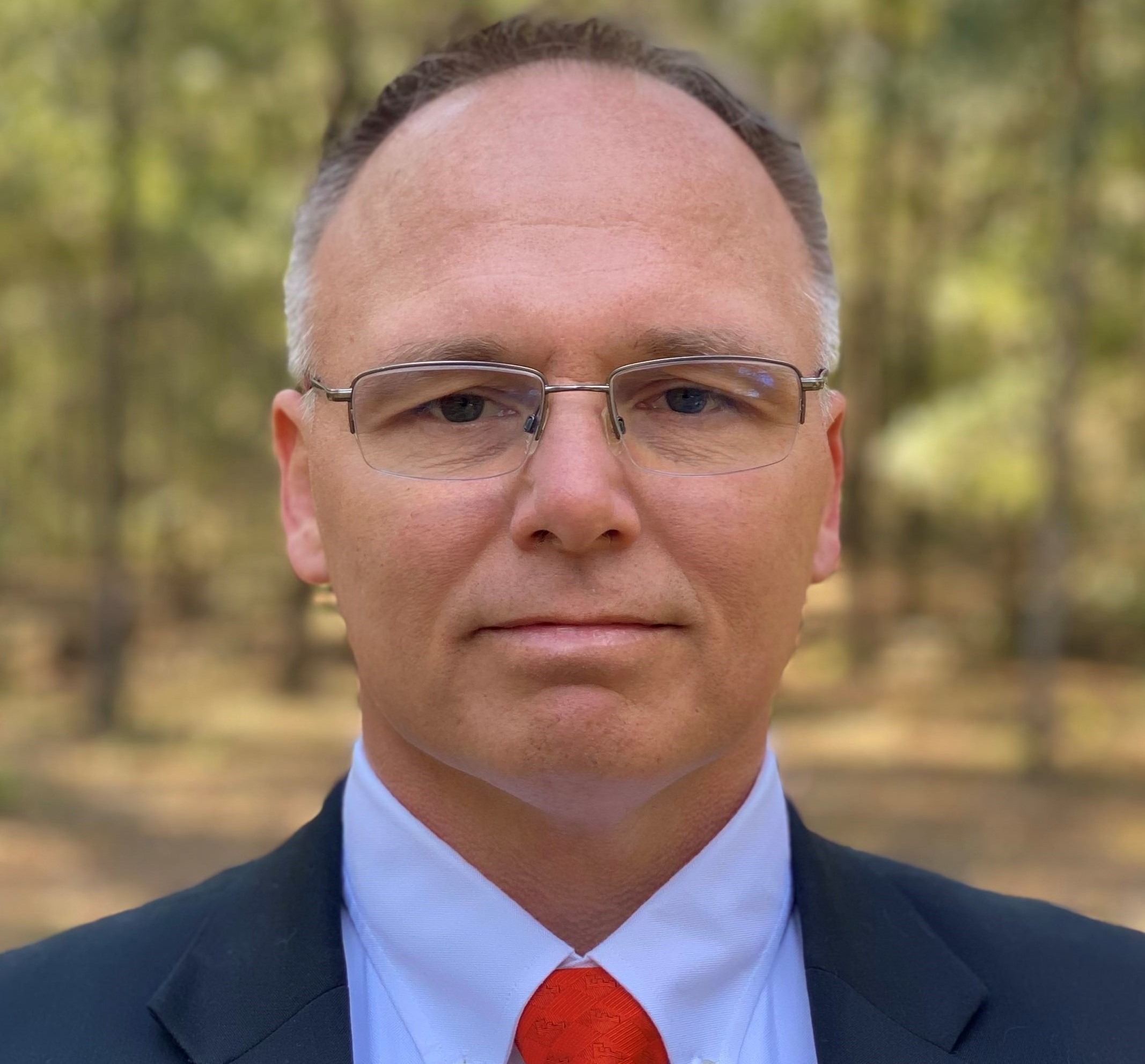}}]{Prof. Warren Dixon}
 received his Ph.D. in 2000 from the Department of Electrical and
Computer Engineering from Clemson University. He worked as a research
staff member and Eugene P. Wigner Fellow at Oak Ridge National Laboratory
(ORNL) until 2004, when he joined the University of Florida in the
Mechanical and Aerospace Engineering Department. His main research
interest has been the development and application of Lyapunov-based
control techniques for uncertain nonlinear systems. His work has been
recognized by the 2019 IEEE Control Systems Technology Award, (2017-2018
\& 2012-2013) University of Florida College of Engineering Doctoral
Dissertation Mentoring Award, 2015 \& 2009 American Automatic Control
Council (AACC) O. Hugo Schuck (Best Paper) Award, the 2013 Fred Ellersick
Award for Best Overall MILCOM Paper, the 2011 American Society of
Mechanical Engineers (ASME) Dynamics Systems and Control Division
Outstanding Young Investigator Award, the 2006 IEEE Robotics and Automation
Society (RAS) Early Academic Career Award, an NSF CAREER Award (2006-2011),
the 2004 Department of Energy Outstanding Mentor Award, and the 2001
ORNL Early Career Award for Engineering Achievement. He is an ASME
Fellow (2016) and IEEE Fellow (2016), was an IEEE Control Systems
Society (CSS) Distinguished Lecturer (2013-2018), served as the Director
of Operations for the Executive Committee of the IEEE CSS Board of
Governors (BOG) (2012-2015), and served as an elected member of the
IEEE CSS BOG (2019-2020). His technical contributions and service
to the IEEE CSS were recognizd by the IEEE CSS Distinguished Member
Award (2020). He was awarded the Air Force Commander's Public Service
Award (2016) for his contributions to the U.S. Air Force Science Advisory
Board.
\end{IEEEbiography}

\end{document}